\documentclass[fleqn,10pt]{wlscirep}
\usepackage[utf8]{inputenc}
\usepackage[T1]{fontenc}
\usepackage{amssymb, amsmath}
\usepackage{color}
\usepackage{url}

\title{Enhanced observation time of magneto-optical traps using micro-machined non-evaporable getter pumps}

\author[1,2*]{Rodolphe Boudot}
\author[1,3]{James P. McGilligan}
\author[1]{Kaitlin R. Moore}
\author[1,3]{Vincent Maurice}
\author[1,3]{Gabriela D. Martinez}
\author[1]{Azure Hansen}
\author[4]{Emeric de Clercq}
\author[1]{John Kitching}
\affil[1]{NIST, Time-Frequency Division, 325 Broadway, Boulder, CO, USA}
\affil[2]{FEMTO-ST, CNRS, 26 rue de l'\'epitaphe, 25000 Besancon, France}
\affil[3]{University of Colorado, Department of Physics, Boulder, Colorado, 80309, USA}
\affil[4]{LNE-SYRTE, Observatoire de Paris, Universit\'e PSL, CNRS, Sorbonne Universit\'e, Paris, France}
\affil[*]{rodolphe.boudot@femto-st.fr}

\begin{abstract}
We show that micro-machined non-evaporable getter pumps (NEGs) can extend the time over which laser cooled atoms can be produced in a magneto-optical trap (MOT), in the absence of other vacuum pumping mechanisms. In a first study, we incorporate a silicon-glass microfabricated ultra-high vacuum (UHV) cell with silicon etched NEG cavities and alumino-silicate glass (ASG) windows and demonstrate the observation of a repeatedly-loading MOT over a 10 minute period with a single laser-activated NEG. In a second study, the capacity of passive pumping with laser activated NEG materials is further investigated in a borosilicate glass-blown cuvette cell containing five NEG tablets. In this cell, the MOT remained visible for over 4 days without any external active pumping system. This MOT observation time exceeds the one obtained in the no-NEG scenario by almost five orders of magnitude. The cell scalability and potential vacuum longevity made possible with NEG materials may enable in the future the development of miniaturized cold-atom instruments.
\end{abstract}
\begin{document}
\flushbottom
\maketitle

\thispagestyle{empty}

\section*{Introduction}
Laser cooling \cite{Hansch:1975, Wineland:1975, Chu:1985, Lett:JOSAB:1989} has permitted groundbreaking advances in fundamental and applied physics by greatly reducing the velocity of atoms, giving access to the detection of narrow atomic resonances\cite{Lett:PRL:1988, Dalibard:1989} and making possible the preparation of pure quantum states \cite{Wineland:1989}. The low momentum ensembles available through laser cooling have led to the development of atomic devices and instruments with unrivaled precision and accuracy, including microwave \cite{Guena:2012} and optical \cite{Huntemann:2016, Schioppo:2017, McGrew:2018, Sanner:2019, Oelker:2019} atomic clocks, quantum sensors \cite{Degen:2017}, magnetometers \cite{Mitchell:2010} and inertial sensors based on matter-wave interferometry \cite{Dutta:2016}.\\ 
The workhorse of cold-atom experiments is the magneto-optical trap (MOT) \cite{PhysRevLett.59.2631}, in which a balanced optical radiation force cools atoms and a spatial localization is created by a magnetic field gradient. The MOT is typically created in an actively-pumped glass-blown cell in which a modest alkali density and ultra-high vacuum (UHV)-level are sustained.\\
In recent years, significant efforts have been made to address the scalability of cold-atom instruments \cite{Bongs:Nature:2019, Garrido:2019}, even resulting in the commercialization of compact cold-atom clocks and sensors. Designs for chip-scale cold-atom systems have also been proposed \cite{Rushton:2014} and demonstrated, including novel ways of redirecting laser beams to trap atoms such as the pyramid MOT \cite{Pollock:09, Gill:2019} and grating-MOT (GMOT) \cite{Nshii:2013, McGilligan:SR:2017,Barker:2019}, as well as density regulators \cite{Kang2019,mcgilliganPRApplied} and low-power coils \cite{Saint2018}. Progress has also been recently reported on the development of chip-scale ion pumps \cite{Basu:2016}. However, the high voltages and large magnetic field in the presence of the atomic sample remain unfavourable for compact atomic clocks and precision instruments.\\
Further miniaturisation of the vacuum cell is possible through the combination of passive pumping techniques and a suitable choice of vacuum materials. For example, micro-electro-mechanical-systems (MEMS) vapor cells, comprised of etched silicon frames and anodically bonded glass windows, provide a means to mass production and micro-fabrication of the vacuum apparatus. Such vapor cells \cite{Kitching:APL:2002, Liew:APL:2004, Knappe:Cells:2005, Douahi:2007, Hasegawa:2011, Vicarini:SA:2018} are now a mature technology, reliable and widely used in chip-scale atomic devices \cite{Kitching:2018}, including commercial products \cite{Lutwak, QuSpin}. Recently, such micro-fabricated cells have demonstrated compatibility with laser cooling through the formation of an actively-pumped MOT in a MEMS platform \cite{mcgilligan2020}.
\\
However, in the absence of active pumping, the residual background pressure in chip-scale cells is rapidly degraded by gas permeation through the glass substrates \cite{Dellis:2016}, material out-gassing, and residual impurities generated during the alkali generation and cell bonding processes \cite{Corman:1998}. In 2012, Scherer \textit{et al.} reported the characterization of alkali metal dispensers and NEG pumps in UHV systems for cold-atom sensors \cite{Scherer:2012} and showed that a MOT could be sustained for several hours in a 500 cm$^3$ volume pumped only with NEGs. In other studies, the activation of thin-film \cite{Hasegawa:2013} or pill-type NEGs \cite{Newman:2018} were demonstrated to mitigate the concentration of impurities in hermetically sealed micro-machined vapor cells.
\\
In this paper, a 6-beam MOT, detected first in a MEMS cell with ASG windows and later in a glass-blown borosilicate cell, is used to study the benefit of laser activated NEGs on the MOT observation time and vacuum pressure longevity with purely passive pumping. Key experimental parameters including the number of atoms trapped in the MOT, the Rb vapor pressure and the non-Rb background pressure are routinely monitored. The MOT observation time, defined as the time taken for the MOT to decay to the detection noise-floor level, was measured to increase by 2 orders of magnitude, up to 10 minutes, after activation of a single NEG in the MEMS cell. An additional test, performed in the conventional borosilicate cuvette-cell with 5 similar NEGs, led to the observation of a MOT for more than 4 days in a regime of pure passive-pumping. This MOT observation time is almost five orders of magnitude longer than in the no-NEG scenario. These results are encouraging for the development of UHV MEMS cells compatible with integrated and low-power cold-atom quantum sensors.

\section*{Methods}
\begin{figure}[b!]
\centering
\includegraphics[width=0.95\linewidth]{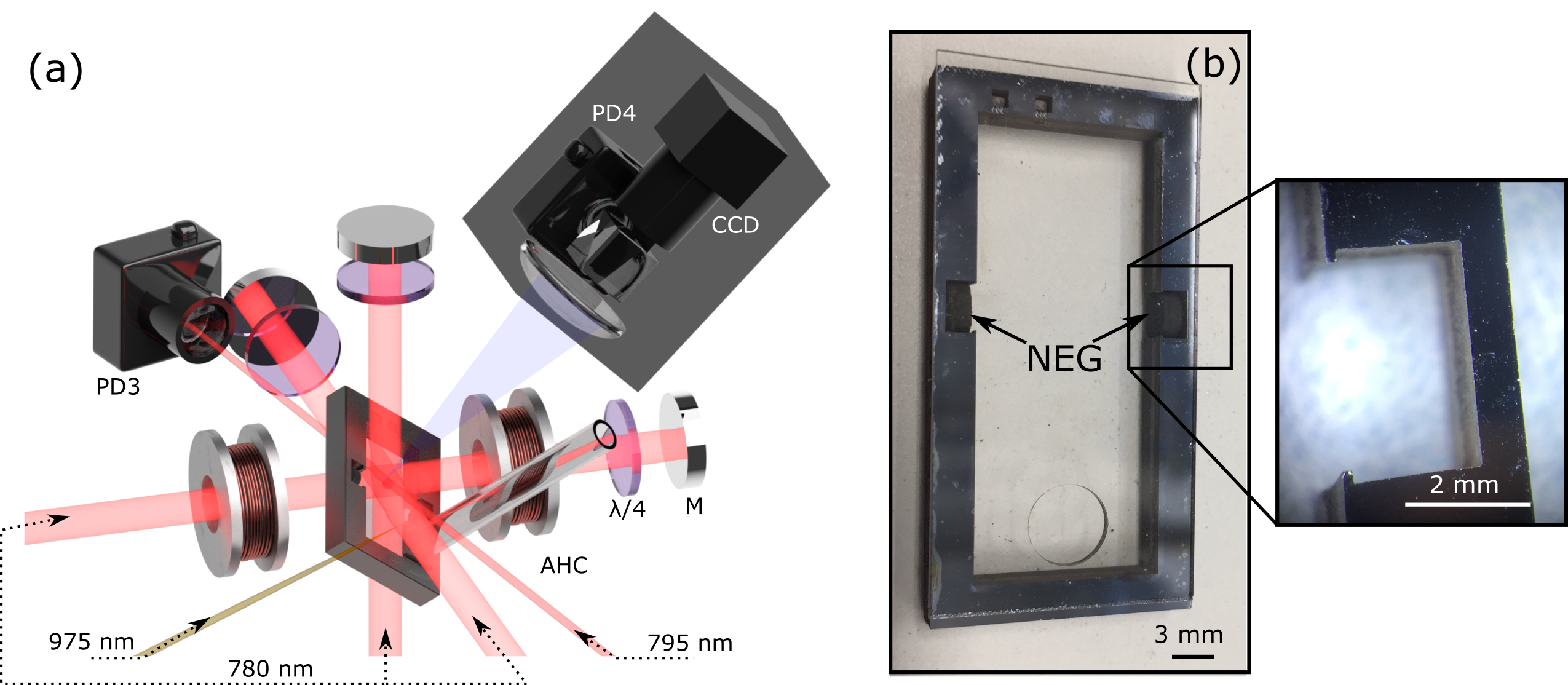}
\caption{(a): Schematic of the MEMS-MOT cell experimental set-up. PD: Photodiode. CCD: Charge-coupled-device. AHC: Anti-Helmholtz coils. M: mirror, (b): MEMS cell after bonding with embedded NEGs. The hole for the vaccum tube connection is visible. A zoom on the NEG cavity is shown.}
\label{fig:Setup}
\end{figure}

\begin{figure}[t]
\centering
\includegraphics[width=0.55\linewidth]{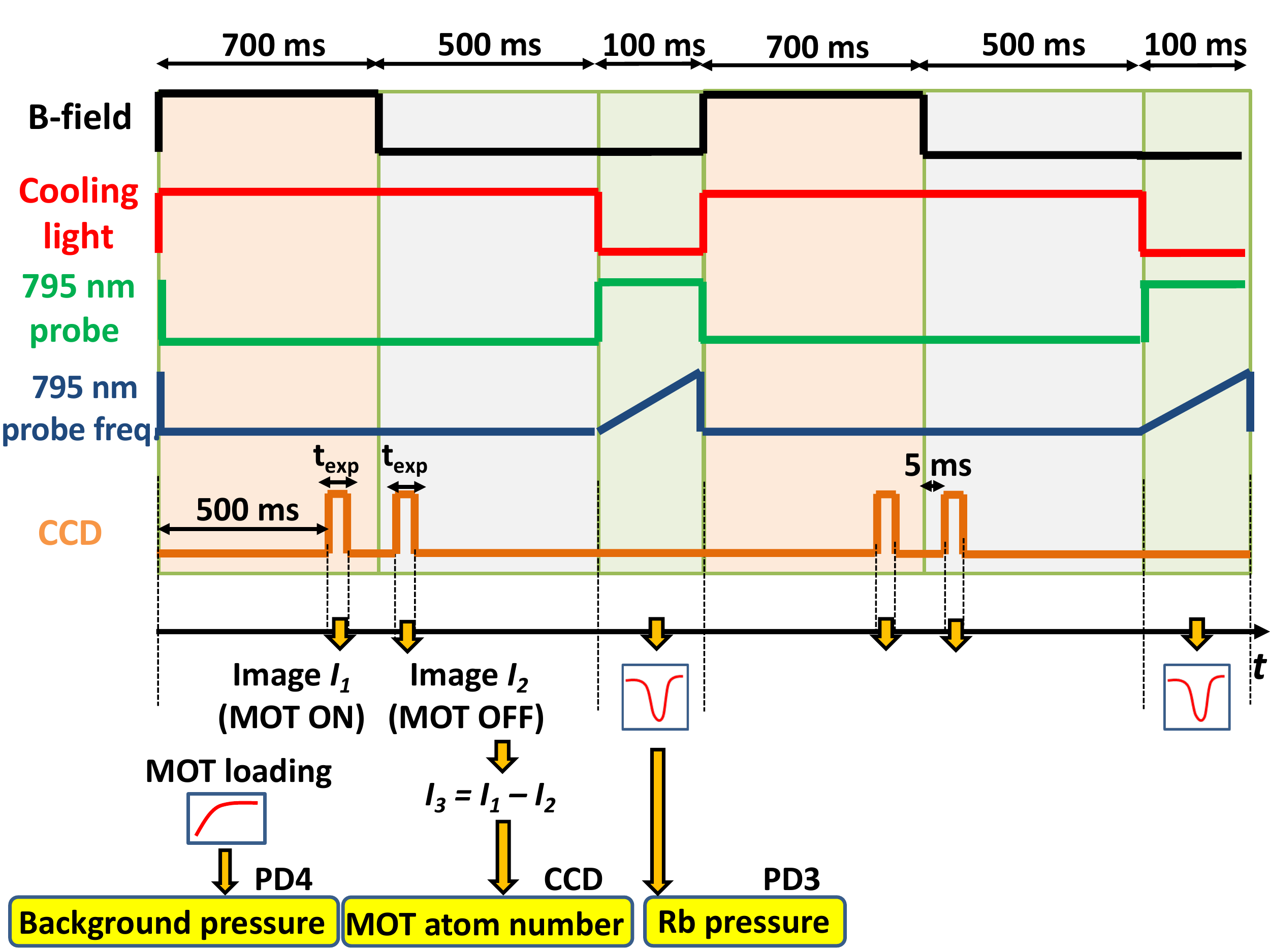}
\caption{Typical sequence of the MEMS-MOT cell measurement. The MOT atom number, the $^{85}$Rb pressure (or density) and the non-Rb background  pressure are routinely measured in the cell throughout the NEGs' activation.}
\label{fig:Sequence}
\end{figure}

Figure~\ref{fig:Setup}(a) shows a simplified schematic of the experimental setup. 
At the center of the laser-cooling system is an actively-pumped micro-fabricated cell. The cell consists of a $40\,\textrm{mm}\times20\,\textrm{mm}\times 4\,\textrm{mm}$ silicon frame etched by deep-reactive ion-etching (DRIE) and sandwiched between two $40\,\textrm{mm}\times20\,\textrm{mm}\times 0.7\,\textrm{mm}$ anodically-bonded low helium permeation aluminosilicate glass wafers (ASG-SD2-Hoya \footnote{Product reference is for technical clarity; does not imply endorsement by NIST})\cite{Dellis:2016}. A 6~mm circular hole is cut through one of the glass windows by laser ablation before anodic bonding to the silicon, allowing the MEMS cell to be connected to an external ion pump via a 7~cm long borosilicate tube. A photograph of the cell, prior to attaching the tube, is shown in Fig. \ref{fig:Setup}(b).
\\
Cavities were etched into the walls of the Si frame to embed non-evaporable getters \footnote{SAES Getters, ST172/WHC/4-2/100. Product reference is for technical clarity and does not imply endorsement by NIST.}, as illustrated in Fig.~\ref{fig:Setup} (b). NEGs are inserted manually into the frame prior to anodic bonding and are held in place by thin 200~$\mu$m fingers to ensure mechanical stability. The active pumping vacuum system contains an electrically-heated alkali-metal dispenser that is used to provide the Rb vapor density.
\\
Rubidium atoms ($^{85}$Rb) are cooled inside the cell using up to 20~mW of total laser power, red-detuned from the D$_2$ cycling transition at 780~nm \cite{MetcalfReviewPaper}. The beam diameter is 8~mm and repumping from the $F=2$ ground state is accomplished by frequency modulating the cooling laser at 2.92~GHz to create an optical sideband at the appropriate detuning. The fluorescence from the MOT is collected using an imaging system with a numerical aperture of 0.4, and imaged onto a CCD. We reduced imaging to the region of interest to mitigate thermal vapor contribution to the MOT counts. A second fluorescence imaging arm connected to a photodiode enables MOT loading time measurements.
\\
NEGs are externally activated by heating with a 1~mm-diameter 975~nm laser beam. During activation of each NEG, the activation laser power was gradually increased until the short term pumping of an individual NEG reached a maximum. A photodiode detects a small amount of light from the activation laser and is used to time stamp the laser activation windows. \\
The measurement sequence is shown in Fig. \ref{fig:Sequence}. An image $I_1$ from the MOT is acquired over a given exposure time (5 or 10 ms) in the presence of both cooling light and magnetic field gradient. The field gradient is then switched off, a background image $I_2$ is taken and a background-subtracted image $I_3= I_1 - I_2$ is then generated. In this sequence, since the cooling laser is ON when the B-field is OFF, the number of counts of the image $I_3$ is actually proportional to $(N_\textrm{MOT} + N_\textrm{hot}) - (N_\textrm{mol} + N_\textrm{hot}) = N_\textrm{MOT} - N_\textrm{mol}$, where $N_\textrm{MOT}$, $N_\textrm{mol}$ and $N_\textrm{hot}$ are the number of atoms actually trapped in the MOT, the number of atoms slowed down by the optical molasses and the number of room-temperature atoms in the vapor, respectively. In our experimental conditions, we calculated using a simplified 1-D model \cite{Lindquist:1992} that $N_\textrm{MOT}/N_\textrm{mol} \simeq 2$. Taking this factor of 2 into account, the MOT atom number was estimated from the number of counts contained in the image $I_3$ using the formula reported in Ref. \cite{Steck:Rb87}.
\\
The residual background pressure in the cell was routinely extracted from measurements of the MOT atom number loading curve time constant using the photodiode fluorescence channel (PD4 on Fig. 1). As shown in Fig. 2, the MOT loading curve is acquired each time the B-field is turned ON (to turn on the MOT). Data points of the MOT atom number loading curve are then approximated by an exponential function of time constant $\tau_{MOT}$ used to calculate the background pressure \cite{Eckel-Tiesinga, Martin:2019}. All background pressure data points shown in figures of this manuscript were obtained using this approach. We mention also that measurements of the background pressure through MOT loading curves were confirmed by measurements of the background pressure extracted from the ion pump current (in situations where the ion pump was activated).
The alkali density probe at 795~nm is aligned through the cell, with the transmission actively monitored on a photodiode (PD3). The density probe is scanned over a GHz range to resolve absorption spectrum within the cell vapor. A lock-in amplifier is used to aid density extraction due to the small absorption path length in the MEMS cell. 

\section*{Results}
\label{sec:Results}
\begin{figure}[t]
\centering
\includegraphics[width=0.95\linewidth]{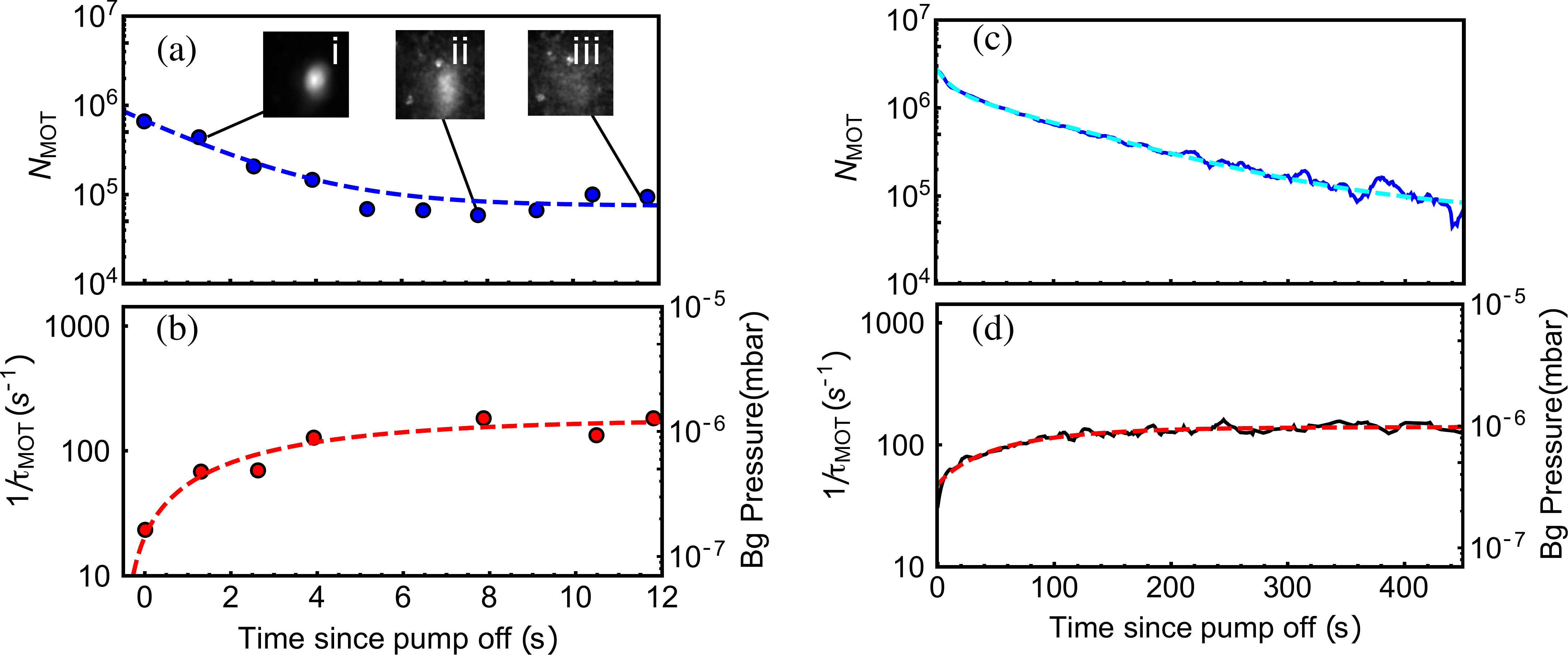}
\caption{(Color online) Decay of the MOT atom-number and evolution of the background pressure versus time in the MEMS-cell in the regime of passive-pumping (after extinction of the ion pump), before (a, b) and after (c, d) the activation of the NEG. Smoothed data (30-pt Savitsky-Golay smoothing) are used for more clarity. In (a), the MOT is not visible after 11~s (no MOT on image (iii)) and data measurement is limited by the imaging system noise-floor. The imaging system noise-floor is higher in (a) than in (c) (after NEG activation), while the initial atom number is smaller, due to an increased cooling power degrading the MOT number and increasing the detected density fluorescence \cite{McGilligan:OE:2015}. Data in (a) are fitted with a single exponential fit function (dashed blue line), with a time constant $\tau_{N}=1.9\pm0.2$ s. In (b), original data points at 5.2, 6.5 and 9.2 s were spurious and were removed. Data shown in (b) are fitted by a single exponential function (red dashed line), with a time constant of $\tau_P=4.2\pm3.5$~s. In (c), after the NEG activation, the evolution of the MOT atom number is found to be well-fitted by a dual-exponential function (cyan dashed line). Background pressure data in (d) are fitted (red dashed line) with the same function as in (b), with an improved time constant of $\tau_P=74\pm9$~s. The very first data points of (b) and (d) each have background pressure levels close to 2~$\times$ 10$^{-7}$~mbar.}
\label{fig:1NEG}
\end{figure}

Prior to NEG activation, a MOT is initially established in the MEMS cell with an electrically driven alkali dispenser. During NEG activation the background pressure and Rb density increase slightly.
Once the atom number again reaches a steady-state, the ion pump is then suddenly turned off and the evolution of the MOT atom number and background pressure are measured, while the Rb density is observed to be constant in the cell. Corresponding results are shown in Fig.~\ref{fig:1NEG}(a) and (b). 
In this configuration, the MOT atom number $N_\textrm{MOT}$ decays rapidly to the detection noise-floor, measured here to be about 8 $\times$ 10$^4$ atoms, within 10~s. Experimental data of the MOT atom number $N_\textrm{MOT}$ are fitted by a single exponential decay function such that $N_\textrm{MOT} (t) = A\times\exp{(-t/\tau_{N})}+c$, with a time constant $\tau_{N}=1.9\pm0.2$ ~s. Simultaneously, the background pressure exponentially increases with the time $t$, following the expected law $P(t)=P_f-\Delta P \exp{(-t/\tau_{P})}$, where $P_f$ is the final pressure, $\Delta P = P_f - P_i$, $P_i$ is the initial pressure, and $\tau_{P}=4.2\pm3.5$~s is the time constant.\\
Following this first test, the NEG is activated and the above-described experiment is repeated, 10 minutes after the end of the activation window, with the results shown in Fig. \ref{fig:1NEG} (c) and (d). In this test, the MOT number decays significantly slower, remaining visible for times exceeding 10~minutes. Thus, with this single NEG activation, an improvement of about 100 was reported in the MOT observation time. In this test, contrary to the test performed before NEG activation, we found that the MOT atom number decay could not be fitted by a single time constant exponential decay function, likely due to the simultaneously evolving background pressure and alkali density within the cell during the initiation of passive pumping. In the present case, as shown in Fig. \ref{fig:1NEG} (c), the MOT atom number decay is found to be well-fitted over 450~s by a dual-exponential function such as $N_\textrm{MOT}(t) = A_1 \times \exp (-t/\tau_{N_1}) + A_2 \times \exp (-t/\tau_{N_2}) + c$, with time constants $\tau_{N_1} = 11 \pm 1$~s and $\tau_{N_2} = 109 \pm 2$~s dominating before and after the first 10~s respectively. This approximation is reported as a phenomenological model. Further studies are required to understand better and model the MOT atom number dynamics. Background pressure data reported in Fig. \ref{fig:1NEG} (d) are again correctly fitted by the expected pressure-rise law, with a time constant $\tau_P=74\pm9$~s. This increased time constant of the vacuum pressure is directly related to the activation of the NEG pump.

\begin{figure}[t]
\centering
\includegraphics[width=0.9\linewidth]{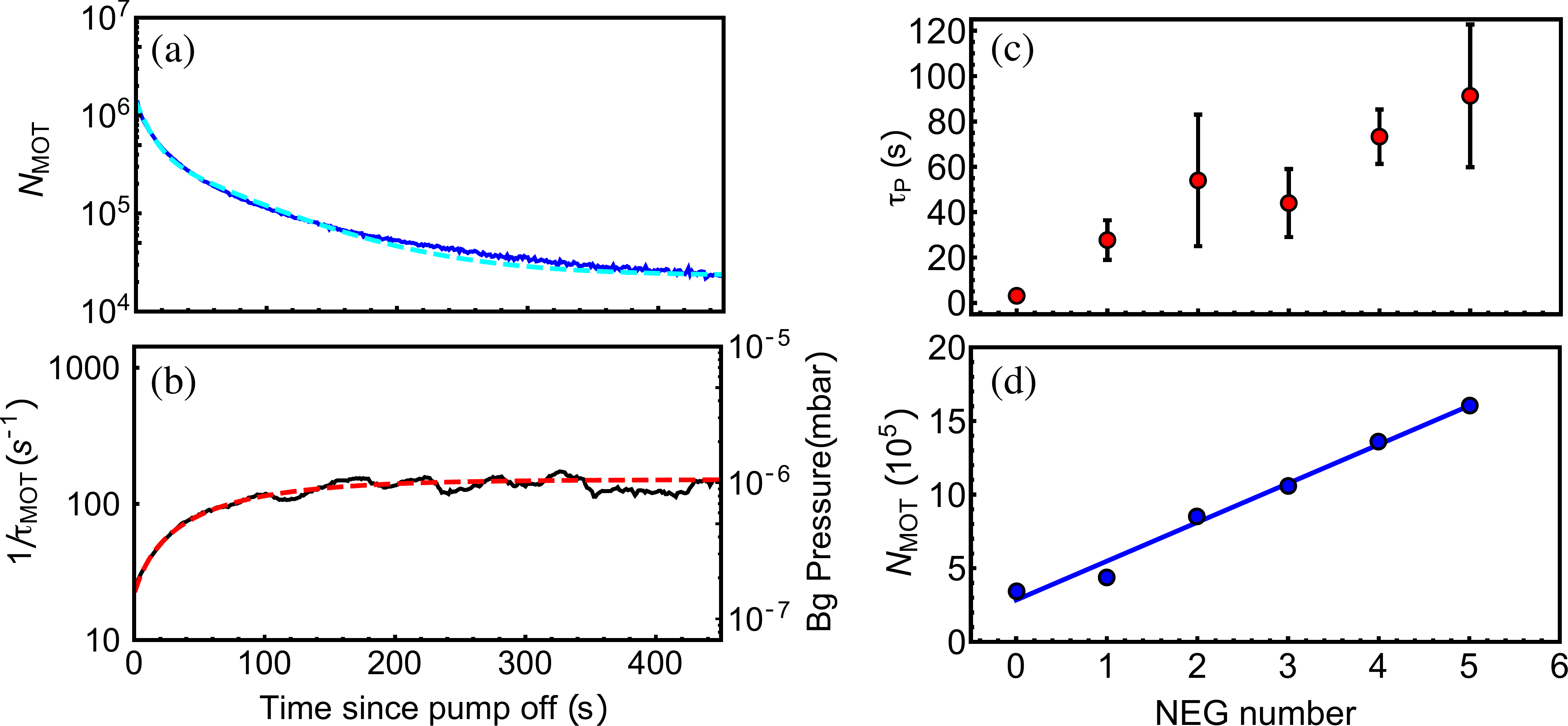}
\caption{(Color online) Decay of the MOT atom number (a) and evolution of the background pressure (b), in the glass-blown cell, after activation of the fourth NEG. In (a), after the NEG activation, the evolution of the MOT atom number is fitted by a dual-exponential function (cyan dashed line). Background pressure data in (b) are fitted by a single exponential function, described in the text, with a time constant $\tau_P =73\pm12$~ s. (c): Values of the time constant $\tau_P$ extracted from the background pressure rise exponential fit versus the number of activated NEGs. Error bars are extracted from the exponential fit, applied to raw (non-smoothed) data of background pressure rise curves. (d): Initial values of the MOT atom number at the beginning of each subsequent NEG activation process, with a linear fit shown as a solid blue line. The cell inner atmosphere is improved after each new activated NEG.}
\label{fig:new-MultiNegs}
\end{figure}

\begin{figure}[t]
\centering
\includegraphics[width=\linewidth]{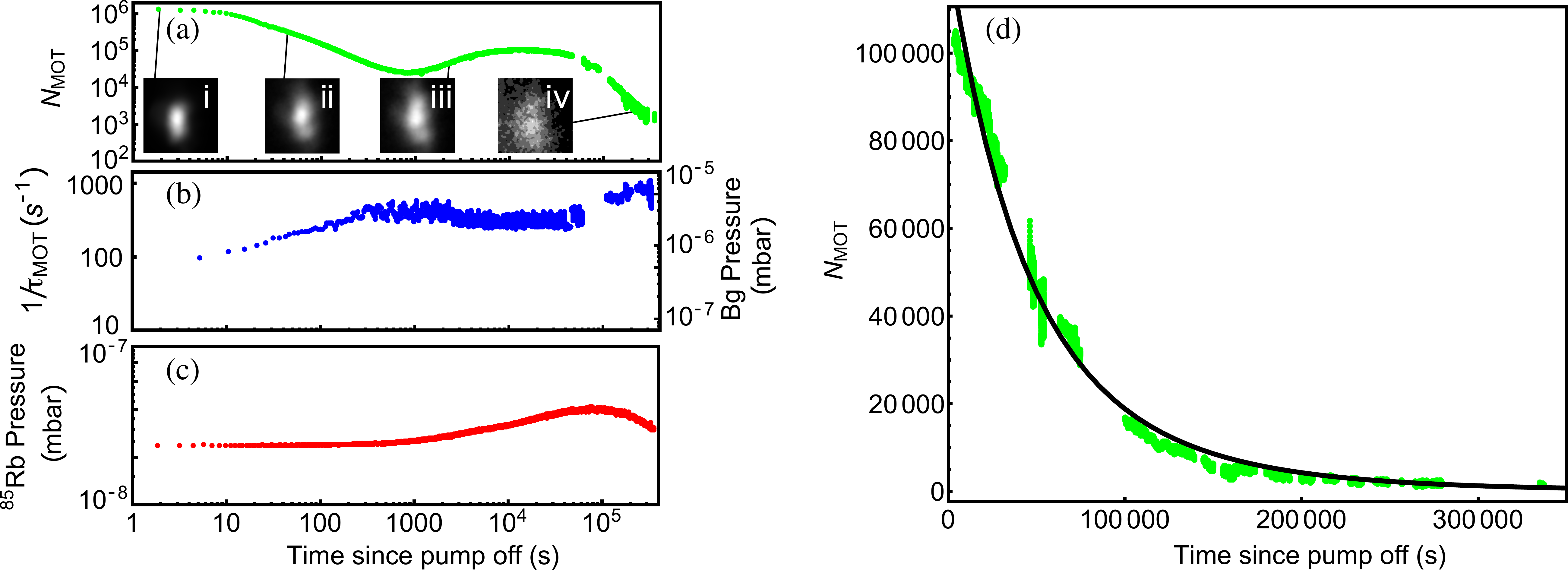}
\caption{(Color online) (a): Steady-state MOT atom number with inset fluorescence images of the MOT at specific times. (b): Background pressure extracted from the MOT loading time constant. (c): $^{85}$Rb pressure for the full time set of the data. (d): MOT atom number evolution extracted from (a) after the bump appearing at about 10$^4$~s. The y-axis is here in linear scale. Experimental data on (d) are fitted by an exponential decay function (solid black line), with a time constant of about 5.2 $\times$ 10$^4$~s. The electric dispenser remained on at a low level throughout the sequence. A 5~s data averaging has been applied to the background pressure data in (b). The absence of points between 6 $\times$ 10$^4$ and 10$^5$~s in (b) is due to a software issue with the MOT loading time extraction. We note on (a) that at long time scales, the MOT exhibits a diffuse shape due to the high background pressure and alkali density regime. In addition, the MOT height likely changes a little over the 4-day measurement. These changes might result from slight mechanical, polarization or optical alignment changes.}
\label{fig:5NEGab}
\end{figure}

\noindent Following the initial demonstration in the MEMS cell, we performed a similar experiment in a standard borosilicate glass-blown cell with a length of 10~cm and a cross-sectional area of 1~cm$^2$, containing 5 NEGs. The use of a glass-blown cell here permits insight to further NEG characterization, without implying the fabrication of a MEMS cell with additional NEG cavities.
The NEGs were sequentially activated while the steady-state MOT atom number and background pressure, as established from the MOT loading curves, were tracked in the absence of active pumping.
In these measurements, the ion pump was turned off 10~minutes after NEG activation and was turned back on again after each new NEG measurement, to let the system reach a new steady-state. Figure \ref{fig:new-MultiNegs} shows an example of the MOT atom number decay (a) and background pressure (b) evolution, following the activation of the fourth NEG and subsequent extinction of the ion pump. In Fig.~\ref{fig:new-MultiNegs}(a), we found that the decay of $N_\textrm{MOT}$ was reasonably fitted by a dual-exponential function, as described above, with $\tau_{N_1} = 10 \pm 0.1$~s and $\tau_{N_2} = 70 \pm 1$~s dominating before and after the first 40~s respectively. The background pressure data (b) are fitted by an exponential function, here with a time constant $\tau_P =73\pm12$~s.
Figure \ref{fig:new-MultiNegs}(c) reports the measured value of the time constant $\tau_P$ versus the number of activated NEGs in the glass-blown cell. We note that the performance of a single activated NEG is less efficient than the single NEG performance in the MEMS cell. This is likely due to the significantly reduced vacuum volume of the MEMS cell compared to the glass-blown cuvette.
It is also observed that the pressure time constant $\tau_P$ increases with each additional NEG activation, showing a summing contribution to the passive pumping within the cell environment. Following the activation of five NEGs, the time constant $\tau_P$ is found to be improved by a factor of 30, in comparison to the initial test (before any NEG activation).
After each passive pumping period with the NEGs, the ion pump was turned back on and the MOT was recovered. We found that with each subsequent NEG activation, the number of atoms in the MOT after ion pump turn-on also increased, as shown in Fig.~\ref{fig:new-MultiNegs} (d). This increase is roughly linear with the number of activated NEGs, showing that the NEGs contribute to the pumping dynamics in the cell in the presence of the ion pump. We noted also that the Rb pressure in the cell increased slightly after each subsequently activated NEG. This could be explained by Rb adsorption onto NEGs prior to activation.\\ 
Following the evaluation of the short-term impact of passive pumping on the cell environment evolution, the mid to long-term evolution of the cell in the regime of purely passive pumping was investigated. Figure~\ref{fig:5NEGab} shows the long-term evolution of the MOT atom number (a), the background pressure (b) and the $^{85}$Rb pressure (c), in the borosilicate cell with 5 activated NEGs, after the ion pump is turned off (at $t=0$).
The MOT atom number, initially at the level of about 10$^6$, is observed to decrease until 1000~s at a value of a few 10$^4$. Following this initial decay, the atom number increases again, flattening around 10$^4$~s before decaying to $2\times10^3$ at $2\times10^5$~s. The resurgence of the MOT number is likely due to a simultaneous decrease of the background pressure and slight increase of the Rb density at 10$^4$~s. The Rb pressure increase can be explained by the fact that the electric dispenser was operating at a fixed current throughout this sequence, leading to a slow increase in alkali vapor pressure due to the lack of active pumping that would otherwise remove Rb from the vacuum. The gradual increase reaches a maximum at 10$^5$~s, where the Rb pressure decreases again until the MOT drops below the detection noise-floor. The reason for the background pressure fluctuation between 10$^3$-10$^4$~s was further investigated. We found that the short-term $\tau_P$ was increased by a factor 10 when a valve was used to remove the ion pump rather than turning it off. This indicates that turning off the ion pump may release contaminants into the vacuum that could take the time scale seen in Fig.~\ref{fig:5NEGab} (b) for the NEG to remove them, resulting in the background pressure fluctuation that is observed.\\
After the resurgence of the MOT atom number near 10$^4$ s, the MOT number decay is fitted by an exponential decay function, shown in Fig.~\ref{fig:5NEGab} (d), with a time constant of $5.2 \times 10^4$~s. The MOT was still clearly visible after 3.5$\times$10$^5$~s, i.e. more than 4 days. Using expressions reported in Ref. \cite{Dellis:2016}, we calculated that He permeation through the borosilicate glass may contribute to the background gas increase at this stage of the experiment. 
We checked that actual variations of the Rb density, cell temperature, magnetic field gradient, total laser intensity or laser detuning, measured during the test, could not explain the MOT atom number dynamics on long integration times. Possible variations of the MOT beam alignment or the MOT beams power distribution (not measured in the experiment) could have contributed to slow variations of the MOT number seen at long observation times \cite{Lindquist:1992}. Although further work is required to demonstrate longer passive pumping times, this proof-of-principle measurement with the activation of five NEGs has yielded the demonstration of a MOT observation time that exceeds 5 orders of magnitude from the no-NEG scenario.\\
\noindent In a last test, to demonstrate that a degradation of the NEGs pumping-rate was not a systematic limitation, the ion pump was re-activated to recover the MOT, before being shut-down again to evaluate the continued pumping performance of the NEGs.
In this scenario, we found that 8 days after the NEGs activation, the values of the time constant $\tau_P$ did not demonstrate any clear sign of degradation of the short-term pumping rate. This result is an additional source of encouragement for the future development of passively-pumped cold-atom MEMS cells.

\section*{Conclusions}
We have reported the detection of a 6-beam magneto-optical trap in a MEMS cell and in a glass-blown cell, each embedding laser-activated passive non-evaporable getter (NEGs) pumps. In each cell, the evolution of the cell inner atmosphere was monitored after achievement of a steady-state MOT thoughout the NEG activation windows and passive pumping tests were later performed by turning off the external active ion pump. In the MEMS cell using ASG windows, a single NEG was successfully laser-activated, demonstrating 2 orders of magnitude improvement of the MOT observation time to 10 minutes. In the glass-blown borosilicate cuvette cell, activation of 5 NEGs yielded a MOT observation time greater than 4 days in the regime of purely passive-pumping, i.e. about five orders of magnitude longer than in the no-NEG scenario. These results open the way to the development of UHV MEMS cells devoted to be exploited in fully-miniaturized cold-atom sensors and instruments.

\bibliography{sample}

\begin{thebibliography}{10}
\urlstyle{rm}
\expandafter\ifx\csname url\endcsname\relax
  \def\url#1{\texttt{#1}}\fi
\expandafter\ifx\csname urlprefix\endcsname\relax\def\urlprefix{URL }\fi
\expandafter\ifx\csname doiprefix\endcsname\relax\def\doiprefix{DOI: }\fi
\providecommand{\bibinfo}[2]{#2}
\providecommand{\eprint}[2][]{\url{#2}}

\bibitem{Hansch:1975}
\bibinfo{author}{Hansch, T.~W.} \& \bibinfo{author}{Schawlow, A.}
\newblock \bibinfo{journal}{\bibinfo{title}{Cooling of gases by laser
  radiation}}.
\newblock {\emph{\JournalTitle{Opt. Comm.}}} \textbf{\bibinfo{volume}{13}},
  \bibinfo{pages}{68--69},
  \doiprefix\url{https://doi.org/10.1016/0030-4018(75)90159-5}
  (\bibinfo{year}{1975}).

\bibitem{Wineland:1975}
\bibinfo{author}{Wineland, D.~J.} \& \bibinfo{author}{Dehmelt, H.}
\newblock \bibinfo{journal}{\bibinfo{title}{Proposed 10$^{14}$ $\delta$ $\nu <
  \nu$ laser fluorescence spectroscopy on tl$^+$ mono-ion oscillator}}.
\newblock {\emph{\JournalTitle{Bulletin of the American Physical Society}}}
  \textbf{\bibinfo{volume}{20}}, \bibinfo{pages}{637} (\bibinfo{year}{1975}).

\bibitem{Chu:1985}
\bibinfo{author}{Chu, S.}, \bibinfo{author}{Hollberg, L.},
  \bibinfo{author}{Bjorkholm, J.~E.}, \bibinfo{author}{A., C.} \&
  \bibinfo{author}{A., A.}
\newblock \bibinfo{journal}{\bibinfo{title}{Three-dimensional viscous
  confinement and cooling of atoms by resonance radiation pressure}}.
\newblock {\emph{\JournalTitle{Phys. Rev. Lett.}}}
  \textbf{\bibinfo{volume}{55}}, \bibinfo{pages}{48},
  \doiprefix\url{https://doi.org/10.1103/PhysRevLett.55.48}
  (\bibinfo{year}{1985}).

\bibitem{Lett:JOSAB:1989}
\bibinfo{author}{Lett, P.~D.} \emph{et~al.}
\newblock \bibinfo{journal}{\bibinfo{title}{Optical molasses}}.
\newblock {\emph{\JournalTitle{J. Opt. Soc. Am. B}}}
  \textbf{\bibinfo{volume}{6}}, \bibinfo{pages}{2084--2107},
  \doiprefix\url{https://doi.org/10.1364/JOSAB.6.002084}
  (\bibinfo{year}{1989}).

\bibitem{Lett:PRL:1988}
\bibinfo{author}{Lett, P.~D.} \emph{et~al.}
\newblock \bibinfo{journal}{\bibinfo{title}{Observation of atoms laser cooled
  below the doppler limit}}.
\newblock {\emph{\JournalTitle{Phys. Rev. Lett.}}}
  \textbf{\bibinfo{volume}{61}}, \bibinfo{pages}{169},
  \doiprefix\url{https://doi.org/10.1103/PhysRevLett.61.169}
  (\bibinfo{year}{1988}).

\bibitem{Dalibard:1989}
\bibinfo{author}{Dalibard, J.} \& \bibinfo{author}{Cohen-Tannoudji, C.}
\newblock \bibinfo{journal}{\bibinfo{title}{Laser cooling below the doppler
  limit by polarization gradients: simple theoretical model}}.
\newblock {\emph{\JournalTitle{J. Opt. Soc. Am. B}}}
  \textbf{\bibinfo{volume}{6}}, \bibinfo{pages}{2023--2045},
  \doiprefix\url{https://doi.org/10.1364/JOSAB.6.002023}
  (\bibinfo{year}{1989}).

\bibitem{Wineland:1989}
\bibinfo{author}{Diedrich, F.}, \bibinfo{author}{Bergquist, J.~C.},
  \bibinfo{author}{Itano, W.~M.} \& \bibinfo{author}{Wineland, D.~J.}
\newblock \bibinfo{journal}{\bibinfo{title}{Laser cooling to the zero-point
  energy of motion}}.
\newblock {\emph{\JournalTitle{Phys. Rev. Lett.}}}
  \textbf{\bibinfo{volume}{62}}, \bibinfo{pages}{403},
  \doiprefix\url{https://doi.org/10.1103/PhysRevLett.62.403}
  (\bibinfo{year}{1989}).

\bibitem{Guena:2012}
\bibinfo{author}{Gu\'ena, J.} \emph{et~al.}
\newblock \bibinfo{journal}{\bibinfo{title}{Progress in atomic fountains at
  lne-syrte}}.
\newblock {\emph{\JournalTitle{IEEE Trans. Ultrason. Ferroelec. Freq. Contr.}}}
  \textbf{\bibinfo{volume}{59}}, \bibinfo{pages}{391--410},
  \doiprefix\url{https://doi.org/10.1109/TUFFC.2012.2208}
  (\bibinfo{year}{2012}).

\bibitem{Huntemann:2016}
\bibinfo{author}{Huntemann, N.}, \bibinfo{author}{Sanner, C.},
  \bibinfo{author}{Lipphardt, B.}, \bibinfo{author}{Tamm, C.} \&
  \bibinfo{author}{Peik, E.}
\newblock \bibinfo{journal}{\bibinfo{title}{Single-ion atomic clock with 3
  $\times$ 10$^{-18}$ systematic uncertainty}}.
\newblock {\emph{\JournalTitle{Phys. Rev. Lett.}}}
  \textbf{\bibinfo{volume}{116}}, \bibinfo{pages}{063001},
  \doiprefix\url{https://doi.org/10.1103/PhysRevLett.116.063001}
  (\bibinfo{year}{2016}).

\bibitem{Schioppo:2017}
\bibinfo{author}{Schioppo, M.} \emph{et~al.}
\newblock \bibinfo{journal}{\bibinfo{title}{Ultra-stable optical clock with two
  cold-atom ensembles}}.
\newblock {\emph{\JournalTitle{Nat. Photon.}}} \textbf{\bibinfo{volume}{11}},
  \bibinfo{pages}{48--52},
  \doiprefix\url{https://doi.org/10.1038/nphoton.2016.231}
  (\bibinfo{year}{2017}).

\bibitem{McGrew:2018}
\bibinfo{author}{McGrew, W.~F.} \emph{et~al.}
\newblock \bibinfo{journal}{\bibinfo{title}{Atomic clock performance enabling
  geodesy below the centimetre level}}.
\newblock {\emph{\JournalTitle{Nature}}} \textbf{\bibinfo{volume}{564}},
  \bibinfo{pages}{87--90},
  \doiprefix\url{https://doi.org/10.1038/s41586-018-0738-2}
  (\bibinfo{year}{2018}).

\bibitem{Sanner:2019}
\bibinfo{author}{Sanner, C.} \emph{et~al.}
\newblock \bibinfo{journal}{\bibinfo{title}{Optical clock comparison for
  lorentz symmetry testing}}.
\newblock {\emph{\JournalTitle{Nature}}} \textbf{\bibinfo{volume}{567}},
  \bibinfo{pages}{204--209},
  \doiprefix\url{https://doi.org/10.1038/s41586-019-0972-2}
  (\bibinfo{year}{2019}).

\bibitem{Oelker:2019}
\bibinfo{author}{Oelker, E.} \emph{et~al.}
\newblock \bibinfo{journal}{\bibinfo{title}{Demonstration of 4.8 $\times$
  10$^{-17}$ stability at 1 s for two independent optical clocks}}.
\newblock {\emph{\JournalTitle{Nat. Photon.}}} \textbf{\bibinfo{volume}{13}},
  \bibinfo{pages}{714--719},
  \doiprefix\url{https://doi.org/10.1038/s41566-019-0493-4}
  (\bibinfo{year}{2019}).

\bibitem{Degen:2017}
\bibinfo{author}{Degen, C.~L.}, \bibinfo{author}{Reinhard, F.} \&
  \bibinfo{author}{Cappellaro, P.}
\newblock \bibinfo{journal}{\bibinfo{title}{Quantum sensing}}.
\newblock {\emph{\JournalTitle{Rev. Mod. Phys.}}}
  \textbf{\bibinfo{volume}{89}}, \bibinfo{pages}{035002},
  \doiprefix\url{https://doi.org/10.1103/RevModPhys.89.035002}
  (\bibinfo{year}{2017}).

\bibitem{Mitchell:2010}
\bibinfo{author}{Koscjorreck, M.}, \bibinfo{author}{Napolitano, M.},
  \bibinfo{author}{Dubost, B.} \& \bibinfo{author}{Mitchell, M.}
\newblock \bibinfo{journal}{\bibinfo{title}{Sub-projection noise sensitivity in
  broadband atomic magnetometry}}.
\newblock {\emph{\JournalTitle{Phys. Rev. Lett.}}}
  \textbf{\bibinfo{volume}{104}}, \bibinfo{pages}{093602},
  \doiprefix\url{https://doi.org/10.1103/PhysRevLett.104.093602}
  (\bibinfo{year}{2010}).

\bibitem{Dutta:2016}
\bibinfo{author}{Dutta, I.} \emph{et~al.}
\newblock \bibinfo{journal}{\bibinfo{title}{Continuous cold-atom inertial
  sensor with 1 nrad/s rotation stability}}.
\newblock {\emph{\JournalTitle{Phys. Rev. Lett.}}}
  \textbf{\bibinfo{volume}{116}}, \bibinfo{pages}{183003},
  \doiprefix\url{https://doi.org/10.1103/PhysRevLett.116.183003}
  (\bibinfo{year}{2016}).

\bibitem{PhysRevLett.59.2631}
\bibinfo{author}{Raab, E.~L.}, \bibinfo{author}{Prentiss, M.},
  \bibinfo{author}{Cable, A.}, \bibinfo{author}{Chu, S.} \&
  \bibinfo{author}{Pritchard, D.~E.}
\newblock \bibinfo{journal}{\bibinfo{title}{Trapping of neutral sodium atoms
  with radiation pressure}}.
\newblock {\emph{\JournalTitle{Phys. Rev. Lett.}}}
  \textbf{\bibinfo{volume}{59}}, \bibinfo{pages}{2631--2634},
  \doiprefix\url{10.1103/PhysRevLett.59.2631} (\bibinfo{year}{1987}).

\bibitem{Bongs:Nature:2019}
\bibinfo{author}{Bongs, K.} \emph{et~al.}
\newblock \bibinfo{journal}{\bibinfo{title}{Taking atom interferometric quantum
  sensors from the laboratory to real- world applications}}.
\newblock {\emph{\JournalTitle{Nature Reviews Physics}}}
  \textbf{\bibinfo{volume}{1}}, \bibinfo{pages}{731--739},
  \doiprefix\url{https://doi.org/10.1038/s42254-019-0117-4}
  (\bibinfo{year}{2019}).

\bibitem{Garrido:2019}
\bibinfo{author}{Garrido~Alzar, C.~L.}
\newblock \bibinfo{journal}{\bibinfo{title}{Compact chip-scale guided cold atom
  gyrometers for inertial navigation: Enabling technologies and design study}}.
\newblock {\emph{\JournalTitle{AVS Quantum Sci.}}}
  \textbf{\bibinfo{volume}{1}}, \bibinfo{pages}{014702},
  \doiprefix\url{https://doi.org/10.1116/1.5120348} (\bibinfo{year}{2019}).

\bibitem{Rushton:2014}
\bibinfo{author}{Rushton, J.}, \bibinfo{author}{Aldous, M.} \&
  \bibinfo{author}{Himsworth, M.}
\newblock \bibinfo{journal}{\bibinfo{title}{Contributed review: The feasability
  of a fully miniaturized magneto-optical trap for portable ultracold quantum
  technology}}.
\newblock {\emph{\JournalTitle{Rev. Sci. Instr.}}}
  \textbf{\bibinfo{volume}{85}}, \bibinfo{pages}{121501},
  \doiprefix\url{http://dx.doi.org/10.1063/1.4904066} (\bibinfo{year}{2014}).

\bibitem{Pollock:09}
\bibinfo{author}{Pollock, S.}, \bibinfo{author}{Cotter, J.~P.},
  \bibinfo{author}{Laliotis, A.} \& \bibinfo{author}{Hinds, E.~A.}
\newblock \bibinfo{journal}{\bibinfo{title}{Integrated magneto-optical traps on
  a silicon pyramid structure}}.
\newblock {\emph{\JournalTitle{Opt. Express}}} \textbf{\bibinfo{volume}{17}},
  \bibinfo{pages}{14109--14114}, \doiprefix\url{10.1364/OE.17.014109}
  (\bibinfo{year}{2009}).

\bibitem{Gill:2019}
\bibinfo{author}{Bowden, W.} \emph{et~al.}
\newblock \bibinfo{journal}{\bibinfo{title}{A pyramid mot with integrated
  optical cavities as a cold atom platform for an optical lattice clock}}.
\newblock {\emph{\JournalTitle{Sci. Rep.}}} \textbf{\bibinfo{volume}{9}},
  \bibinfo{pages}{11704},
  \doiprefix\url{https://doi.org/10.1038/s41598-019-48168-3}
  (\bibinfo{year}{2019}).

\bibitem{Nshii:2013}
\bibinfo{author}{Nshii, C.~C.} \emph{et~al.}
\newblock \bibinfo{journal}{\bibinfo{title}{A surface-patterned chip as a
  strong source of ultracold atoms for quantum technologies}}.
\newblock {\emph{\JournalTitle{Nature Nanotech.}}}
  \textbf{\bibinfo{volume}{8}}, \bibinfo{pages}{321--324},
  \doiprefix\url{https://doi.org/10.1038/nnano.2013.47} (\bibinfo{year}{2013}).

\bibitem{McGilligan:SR:2017}
\bibinfo{author}{McGilligan, J.~P.} \emph{et~al.}
\newblock \bibinfo{journal}{\bibinfo{title}{Grating chips for quantum
  technologies}}.
\newblock {\emph{\JournalTitle{Sci. Rep.}}} \textbf{\bibinfo{volume}{7}},
  \bibinfo{pages}{384},
  \doiprefix\url{https://doi.org/10.1038/s41598-017-00254-0}
  (\bibinfo{year}{2017}).

\bibitem{Barker:2019}
\bibinfo{author}{Barker, D.~S.} \emph{et~al.}
\newblock \bibinfo{journal}{\bibinfo{title}{Single-beam zeeman slower and
  magneto-optical trap using a nanofabricated grating}}.
\newblock {\emph{\JournalTitle{Phys. Rev. Applied}}}
  \textbf{\bibinfo{volume}{11}}, \bibinfo{pages}{064023},
  \doiprefix\url{https://doi.org/10.1103/PhysRevApplied.11.064023}
  (\bibinfo{year}{2019}).

\bibitem{Kang2019}
\bibinfo{author}{Kang, S.} \emph{et~al.}
\newblock \bibinfo{journal}{\bibinfo{title}{Magneto-optic trap using a
  reversible, solid-state alkali-metal source}}.
\newblock {\emph{\JournalTitle{Optics Letters}}} \textbf{\bibinfo{volume}{44}},
  \bibinfo{pages}{3002}, \doiprefix\url{https://doi.org/10.1364/ol.44.003002}
  (\bibinfo{year}{2019}).

\bibitem{mcgilliganPRApplied}
\bibinfo{author}{McGilligan, J.~P.} \emph{et~al.}
\newblock \bibinfo{journal}{\bibinfo{title}{Dynamic characterization of an
  alkali-ion battery as a source for laser-cooled atoms}}.
\newblock {\emph{\JournalTitle{Physical Review Applied}}}
  \textbf{\bibinfo{volume}{13}}, \bibinfo{pages}{044038--},
  \doiprefix\url{10.1103/PhysRevApplied.13.044038} (\bibinfo{year}{2020}).

\bibitem{Saint2018}
\bibinfo{author}{Saint, R.} \emph{et~al.}
\newblock \bibinfo{journal}{\bibinfo{title}{3d-printed components for quantum
  devices}}.
\newblock {\emph{\JournalTitle{Scientific Reports}}}
  \textbf{\bibinfo{volume}{8}},
  \doiprefix\url{https://doi.org/10.1038/s41598-018-26455-9}
  (\bibinfo{year}{2018}).

\bibitem{Basu:2016}
\bibinfo{author}{Basu, A.} \& \bibinfo{author}{Velasquez-Garcia, L.~F.}
\newblock \bibinfo{journal}{\bibinfo{title}{An electrostratic ion pump with
  nanostructured si field emission eletron source and ti particle collectors
  for supporting an ultra-high vacuum in miniaturized atom interferometry
  systems}}.
\newblock {\emph{\JournalTitle{J. Micromech. Microeng.}}}
  \textbf{\bibinfo{volume}{26}}, \bibinfo{pages}{124003}
  (\bibinfo{year}{2016}).

\bibitem{Kitching:APL:2002}
\bibinfo{author}{Kitching, J.}, \bibinfo{author}{Knappe, S.} \&
  \bibinfo{author}{Hollberg, L.}
\newblock \bibinfo{journal}{\bibinfo{title}{Miniature vapor-cell
  atomic-frequency references}}.
\newblock {\emph{\JournalTitle{Appl. Phys. Lett.}}}
  \textbf{\bibinfo{volume}{81}}, \bibinfo{pages}{553--555},
  \doiprefix\url{https://doi.org/10.1063/1.1494115} (\bibinfo{year}{2002}).

\bibitem{Liew:APL:2004}
\bibinfo{author}{Liew, L.} \emph{et~al.}
\newblock \bibinfo{journal}{\bibinfo{title}{Microfabricated alkali atom vapor
  cells}}.
\newblock {\emph{\JournalTitle{Appl. Phys. Lett.}}}
  \textbf{\bibinfo{volume}{84}}, \bibinfo{pages}{2694--2696},
  \doiprefix\url{https://doi.org/10.1063/1.1691490} (\bibinfo{year}{2004}).

\bibitem{Knappe:Cells:2005}
\bibinfo{author}{Knappe, S.} \emph{et~al.}
\newblock \bibinfo{journal}{\bibinfo{title}{Atomic vapor cells for chip-scale
  atomic clocks with improved long-term frequency stability}}.
\newblock {\emph{\JournalTitle{Opt. Lett.}}} \textbf{\bibinfo{volume}{30}},
  \bibinfo{pages}{2351--2353},
  \doiprefix\url{https://doi.org/10.1364/OL.30.002351} (\bibinfo{year}{2005}).

\bibitem{Douahi:2007}
\bibinfo{author}{Douahi, A.} \emph{et~al.}
\newblock \bibinfo{journal}{\bibinfo{title}{Vapour microcell for chip scale
  atomic frequency standard}}.
\newblock {\emph{\JournalTitle{Elec. Lett.}}} \textbf{\bibinfo{volume}{43}},
  \bibinfo{pages}{33--34}, \doiprefix\url{https://doi.org/10.1049/el:20070147}
  (\bibinfo{year}{2007}).

\bibitem{Hasegawa:2011}
\bibinfo{author}{Hasegawa, M.} \emph{et~al.}
\newblock \bibinfo{journal}{\bibinfo{title}{Microfabrication of cesium vapor
  cells with buffer gas for mems atomic clocks}}.
\newblock {\emph{\JournalTitle{Sensors Actuators: Phys. A}}}
  \textbf{\bibinfo{volume}{167}}, \bibinfo{pages}{594--601},
  \doiprefix\url{https://doi.org/10.1016/j.sna.2011.02.039}
  (\bibinfo{year}{2011}).

\bibitem{Vicarini:SA:2018}
\bibinfo{author}{Vicarini, R.} \emph{et~al.}
\newblock \bibinfo{journal}{\bibinfo{title}{Demonstration of the
  mass-producible feature of a cs vapor microcell technology for miniature
  atomic clocks}}.
\newblock {\emph{\JournalTitle{Sensors Actuators: Phys. A}}}
  \textbf{\bibinfo{volume}{280}}, \bibinfo{pages}{99--106},
  \doiprefix\url{https://doi.org/10.1016/j.sna.2018.07.032}
  (\bibinfo{year}{2018}).

\bibitem{Kitching:2018}
\bibinfo{author}{Kitching, J.}
\newblock \bibinfo{journal}{\bibinfo{title}{Chip-scale atomic devices}}.
\newblock {\emph{\JournalTitle{Appl. Phys. Rev.}}}
  \textbf{\bibinfo{volume}{5}}, \bibinfo{pages}{031302},
  \doiprefix\url{https://doi.org/10.1063/1.5026238} (\bibinfo{year}{2018}).

\bibitem{Lutwak}
\bibinfo{author}{Lutwak, R.} \emph{et~al.}
\newblock \bibinfo{journal}{\bibinfo{title}{The miniature atomic clock –
  pre-production results}}.
\newblock {\emph{\JournalTitle{2007 IEEE International Frequency Control
  Symposium Joint with the 21st European Frequency and Time Forum, Geneva,
  Switzerland}}} \bibinfo{pages}{1327--133},
  \doiprefix\url{https://ieeexplore.ieee.org/document/4319292/}
  (\bibinfo{year}{2007}).

\bibitem{QuSpin}
\bibinfo{author}{Shah, V.~K.} \& \bibinfo{author}{Wakai, R.~T.}
\newblock \bibinfo{journal}{\bibinfo{title}{A compact, high performance atomic
  magnetometer for biomedical applications}}.
\newblock {\emph{\JournalTitle{Phys. Med. Biol.}}}
  \textbf{\bibinfo{volume}{58}}, \bibinfo{pages}{8153--8161},
  \doiprefix\url{https://doi.org/10.1088/0031-9155/58/22/8153}
  (\bibinfo{year}{2013}).

\bibitem{mcgilligan2020}
\bibinfo{author}{McGilligan, J.~P.} \emph{et~al.}
\newblock \bibinfo{journal}{\bibinfo{title}{Laser cooling in a chip-scale
  platform}}.
\newblock {\emph{\JournalTitle{Appl. Phys. Lett.}}}
  \textbf{\bibinfo{volume}{117}}, \bibinfo{pages}{054001}
  (\bibinfo{year}{2020}).

\bibitem{Dellis:2016}
\bibinfo{author}{Dellis, A.~T.}, \bibinfo{author}{Shah, V.},
  \bibinfo{author}{Donley, E.~A.}, \bibinfo{author}{Knappe, S.} \&
  \bibinfo{author}{Kitching, J.}
\newblock \bibinfo{journal}{\bibinfo{title}{Low helium permeation cells for
  atomic microsystems technology}}.
\newblock {\emph{\JournalTitle{Opt. Lett.}}} \textbf{\bibinfo{volume}{41}},
  \bibinfo{pages}{2775--2778},
  \doiprefix\url{https://doi.org/10.1364/OL.41.002775} (\bibinfo{year}{2016}).

\bibitem{Corman:1998}
\bibinfo{author}{Corman, T.}, \bibinfo{author}{Enokson, P.} \&
  \bibinfo{author}{Stemme, G.}
\newblock \bibinfo{journal}{\bibinfo{title}{Low-pressure-encapsulated resonant
  structures with integrated electrodes for electrostatic excitation and
  capacitive detection}}.
\newblock {\emph{\JournalTitle{Sensors Actuators: Phys. A}}}
  \textbf{\bibinfo{volume}{66}}, \bibinfo{pages}{160--166},
  \doiprefix\url{https://doi.org/10.1016/S0924-4247(98)80019-8}
  (\bibinfo{year}{1998}).

\bibitem{Scherer:2012}
\bibinfo{author}{Scherer, D.~R.}, \bibinfo{author}{Fenner, D.~B.} \&
  \bibinfo{author}{M., H.~J.}
\newblock \bibinfo{journal}{\bibinfo{title}{Characterization of alkali metal
  dispensers and non-evaporable getter pumps in ultra-high vacuum systems for
  cold atomic sensors}}.
\newblock {\emph{\JournalTitle{J. Vac. Sci. Technol.}}}
  \textbf{\bibinfo{volume}{30}}, \bibinfo{pages}{061602},
  \doiprefix\url{http://dx.doi.org/10.1116/1.4757950} (\bibinfo{year}{2012}).

\bibitem{Hasegawa:2013}
\bibinfo{author}{Hasegawa, M.} \emph{et~al.}
\newblock \bibinfo{journal}{\bibinfo{title}{Effects of getters on hermetically
  sealed micromachined cesium-neon cells for atomic clocks}}.
\newblock {\emph{\JournalTitle{J. Micromech. Microeng.}}}
  \textbf{\bibinfo{volume}{23}}, \bibinfo{pages}{055022},
  \doiprefix\url{https://doi.org/10.1088/0960-1317/23/5/055022}
  (\bibinfo{year}{2013}).

\bibitem{Newman:2018}
\bibinfo{author}{Newman, Z.~L.} \emph{et~al.}
\newblock \bibinfo{journal}{\bibinfo{title}{Architecture for the photonic
  integration of an optical atomic clock}}.
\newblock {\emph{\JournalTitle{Optica}}} \textbf{\bibinfo{volume}{6}},
  \bibinfo{pages}{680--685},
  \doiprefix\url{https://doi.org/10.1364/OPTICA.6.000680}
  (\bibinfo{year}{2018}).

\bibitem{MetcalfReviewPaper}
\bibinfo{author}{Metcalf, H.~J.} \& \bibinfo{author}{van~der Straten, P.}
\newblock \bibinfo{journal}{\bibinfo{title}{Laser cooling and trapping of
  atoms}}.
\newblock {\emph{\JournalTitle{J. Opt. Soc. Am. B}}}
  \textbf{\bibinfo{volume}{20}}, \bibinfo{pages}{887--908},
  \doiprefix\url{https://doi.org/10.1364/JOSAB.20.000887}
  (\bibinfo{year}{2003}).

\bibitem{Lindquist:1992}
\bibinfo{author}{Lindquist, K.}, \bibinfo{author}{Stephens, M.} \&
  \bibinfo{author}{Wieman, C.}
\newblock \bibinfo{journal}{\bibinfo{title}{Experimental and theoretical study
  of the vapor-cell zeeman optical trap}}.
\newblock {\emph{\JournalTitle{Phys. Rev. A}}} \textbf{\bibinfo{volume}{46}},
  \bibinfo{pages}{4082},
  \doiprefix\url{https://doi.org/10.1103/PhysRevA.46.4082}
  (\bibinfo{year}{1992}).

\bibitem{Steck:Rb87}
\bibinfo{author}{Steck, D.~A.}
\newblock \bibinfo{title}{Rb 85 d line data} (\bibinfo{year}{2001}).

\bibitem{Eckel-Tiesinga}
\bibinfo{author}{Eckel, S.} \emph{et~al.}
\newblock \bibinfo{journal}{\bibinfo{title}{Challenges to miniaturizing cold
  atom technology for deployable vacuum metrology}}.
\newblock {\emph{\JournalTitle{Metrologia}}} \textbf{\bibinfo{volume}{55}},
  \bibinfo{pages}{S182},
  \doiprefix\url{https://doi.org/10.1088/1681-7575/aadbe4}
  (\bibinfo{year}{2018}).

\bibitem{Martin:2019}
\bibinfo{author}{Martin, J.~M.} \emph{et~al.}
\newblock \bibinfo{journal}{\bibinfo{title}{Pumping dynamics of cold-atom
  experiments in a single vacuum chamber}}.
\newblock {\emph{\JournalTitle{Phys. Rev. Applied}}}
  \textbf{\bibinfo{volume}{12}}, \bibinfo{pages}{014033},
  \doiprefix\url{https://doi.org/10.1103/PhysRevApplied.12.014033}
  (\bibinfo{year}{2019}).

\bibitem{McGilligan:OE:2015}
\bibinfo{author}{McGilligan, J.~P.}, \bibinfo{author}{Griffin, P.~F.},
  \bibinfo{author}{Riis, E.} \& \bibinfo{author}{Arnold, A.~S.}
\newblock \bibinfo{journal}{\bibinfo{title}{Phase-space properties of
  magneto-optical traps utilising micro-fabricated gratings}}.
\newblock {\emph{\JournalTitle{Opt. Exp.}}} \textbf{\bibinfo{volume}{23}},
  \bibinfo{pages}{8948--8959},
  \doiprefix\url{https://doi.org/10.1364/OE.23.008948} (\bibinfo{year}{2015}).

\end{thebibliography}

\section*{Acknowledgements}
The authors acknowledge Alejandra Collopy, Matt Simons, Elizabeth Donley, William McGehee (NIST), Carlos Garrido Alzar (SYRTE) and Aidan Arnold (University of Strathclyde) for careful reading of the manuscript before submission. The authors thank Y-J. Chen and M. Shuker for fruitful discussions. R.B was supported by the NIST Guest Researcher program and D\'el\'egation G\'en\'erale de l'Armement (DGA). J.P.M. gratefully acknowledges support from the English Speaking Union and Lindemann Fellowship. G. D. M. was supported under the financial assistance award 70NANB18H006 from the U.S. Department of Commerce, National Institute of Standards and Technology. 

\section*{Author contributions statement}
R. B and J. P. M contributed equally to this work. They contributed to the design, fabrication and implementation in the MOT system of the NEG-pill cells, participated to the development of the MOT imaging experiment software, conducted experiments, analyzed results and shared the manuscript writing process. K.M and G.M. contributed to the MEMS cell technology development and to the experimental setup. V.M. contributed to the MEMS cell design and to the development of the MOT imaging and analysis software. A. H. and E. de C. contributed to results analysis and helped with the manuscript writing. J. K. oversaw the project and contributed to the results analysis and writing of the manuscript. All authors reviewed the manuscript. 

\end{document}